\documentclass[12pt]{amsart}
\usepackage[latin1]{inputenc}
\usepackage{latexsym}
\usepackage{amsmath}
\usepackage{amssymb}
\usepackage{amsfonts}
\usepackage{enumerate}
\usepackage{multicol}
 \newtheorem{thm}{Theorem}[section]
 
 \newtheorem{lem}[thm]{Lemma}
 
 \newtheorem{prop}[thm]{Proposition}
 \theoremstyle{definition}
 
 \theoremstyle{remark}

 \numberwithin{equation}{section}
\newcommand{\wt}[1]{\widetilde{#1}}
\newcommand{\SPn}[2]{\langle \,#1\,|\,#2\, \rangle}

\renewcommand{\le}{\leqslant} % greater/less or equal         
\renewcommand{\ge}{\geqslant}

\renewcommand{\imath}{i}
\newcommand{\R}{\mathcal{R}}

\newcommand{\s}{{\rm S}}
\renewcommand{\r}{{\rm R}}

\renewcommand{\S}{{\rm S}}
\renewcommand{\R}{{\rm R}}
\newcommand{\bbbone}{{\mathbf 1}}
\newcommand{\e}{{\rm e}}
\renewcommand{\i}{{\rm i}}
\newcommand{\scalprod}[2]{\left\langle {#1}, {#2}\right\rangle}
\begin{document}

\title{Completely positive dynamical semigroups \\ and\\ quantum resonance theory}

%\author{M. K\"onenberg  {\protect \and} M. Merkli}

\address[Martin K\"onenberg]{Department of Mathematics and Statistics, Memorial University, St. John's, NL, Canada}
\author{Martin K\"onenberg}
\curraddr{Fachbereich Mathematik, Universit\"at Stuttgart, Germany} 
\email{martin.koenenberg@mathematik.uni-stuttgart.de}
\address[Marco Merkli]{Department of Mathematics and Statistics, Memorial University, St. John's, NL, Canada}
\author{Marco Merkli}
\email{merkli@mun.ca}

\date{December 29, 2016}

\begin{abstract}
Starting form a microscopic system-environment model, we construct a quantum dynamical semigroup for the reduced evolution of the open system. The difference between the true system dynamics and its approximation by the semigroup has the following two properties: It is (linearly) small in the system-environment coupling constant for all times, and it vanishes exponentially quickly in the large time limit. Our approach is based on the quantum dynamical resonance theory.
\end{abstract}

\maketitle

\section{The issue}

Due to the entanglement of an open system with its surroundings, its dynamics $V(t): \rho_0\mapsto \rho_t$, mapping an initial system density matrix $\rho_0$ to its value at time $t$, is not a semigroup in time. For each fixed $t$, the mapping $V(t)$, called a {\em dynamical map}, is a linear, completely positive, trace preserving transformation.\footnote{We recall that a map on bounded operators on a Hilbert space,  $V:{\mathcal B}({\mathcal H})\rightarrow {\mathcal B}({\mathcal H})$, is called completely positive if $V\otimes\bbbone:{\mathcal B}({\mathcal H}\otimes{\mathbb C}^n)\rightarrow {\mathcal B}({\mathcal H}\otimes{\mathbb C}^n)$ is positive (maps positive operators into positive ones) for all $n\in\mathbb N$. } Under certain assumptions, one can approximate the dynamics of an open system by a continuous one-parameter semigroup of dynamical maps, called a {\em quantum dynamical semigroup} \cite{BP,GZ}. The dynamics given by such a semigroup has two important features: (i) is it markovian due to the semigroup property and (ii) it maps density matrices into density matrices due to its trace and positivity preserving quality. {\em Complete} positivity of the dynamical semigroup implies its positivity preservation, but not vice-versa. It is a crucial physical property which ensures that the dynamics of initially entangled systems interacting with an environment is well defined \cite{AL,BF}. The semigroup property is particularly convenient since the spectral analysis of the generator $\mathcal L$ of the semigroup yields dynamical properties of the system, such as the final state(s) and convergence speeds. Controlling the remainder in the approximation $V(t)\rho_0 \approx \e^{t{\mathcal L}}\rho_0$ rigorously is difficult. Microscopic derivations, passing from a full (hamiltonian) model of system plus environment and tracing out the environment degrees of freedom, involve approximations (Born, Markov,  rotating wave) that are hard to deal with mathematically. In some situations where the  system-environment interaction is weak, measured by a small coupling constant $\lambda$, one can implement a (time-dependent) perturbation theory, $\lambda=0$ giving the unperturbed (uncoupled) case. For certain systems it has been shown \cite{Davies} that for all $a>0$, 
$$
\lim_{\lambda\rightarrow 0} \sup_{0\le \lambda^2 t<a}\| V(t) - \e^{t({\mathcal L}_0+\lambda^2 K)} \| =0,
$$
where ${\mathcal L}_0$ is the generator of the uncoupled (hamiltonian) dynamics and $K$ is a (lowest order) correction responsible for dissipative effects. We discuss here finite-dimensional open systems and so the nature of the norm is immaterial. This weak coupling result allows a description of the dynamics by a semigroup up to times $t=O(\lambda^{-2})$. However, the asymptotics $t\rightarrow\infty$ is  not resolved correctly by the dynamical semigroup $\e^{t({\mathcal L}_0+\lambda^2 K)}$. For instance, the invariant state of the latter is typically the {\em uncoupled} system Gibbs equilibrium state, while the true asymptotic state is the restriction of the {\em coupled} system-environment equilibrium to the system alone. A more recent dynamical resonance theory \cite{JP1, Metal} improves the weak coupling result to
$$
\|V(t) -\e^{t M(\lambda)}\| \le  C \lambda^2\e^{-\gamma' t}, \qquad t\ge 0,
$$
where $M(\lambda)={\mathcal L}_0+\lambda^2 K+\cdots$ is analytic in $\lambda$ and $\gamma'>0$. This approach grew out of works proving `return to equilibrium' of open systems using elements of algebraic quantum field theory and spectral theory \cite{JP0,BFS} and is useful in different physical settings \cite{Mappli}. While it is known that the `Davies generator' ${\mathcal L}_0+\lambda^2 K$, describing the weak coupling limit, generates a dynamical semigroup \cite{Davies, DS}, this is not known for the generator $M(\lambda)$ emerging from the dynamical resonance theory. In the present paper, we construct a dynamical semigroup ${\mathcal T}_t$ satisfying
$$
 \qquad {\mathcal T}_0=\bbbone,\qquad {\mathcal T}_{t+s} = {\mathcal T}_{t}\,  {\mathcal T}_{s},\quad \forall s,t\ge 0
$$
and
$$
\| V(t) -{\mathcal T}_t\| \le 
C|\lambda|\, \big(1+\lambda^2 t\big)\,\e^{-\lambda^2(1+O(\lambda))\gamma t}, \qquad t \ge 0,
$$
where $\gamma>0$.
Giving the Schr\"odinger dynamics ${\mathcal T}_{t}$ is equivalent to giving the completely positive, identity preserving semigroup $\tau^t$ acting on the algebra of observables of the system (Heisenberg dynamics), defined by the relation ${\rm Tr}( \{{\mathcal T}_{t}\rho_0\} X)= {\rm Tr}(\rho_0\, \tau^t(X))$ for all system density matrices $\rho_0$ and all system observables $X$. Our main result, Theorem \ref{mainthm}, shows the existence of the Heisenberg dynamics $\tau^t$. We construct it  by modifying the dynamical resonance theory approach right at its starting point. Namely, instead of taking the uncoupled system equilibrium state as a reference state, we take for it  the effective, coupled system equilibrium state, which contains all orders of interactions with the reservoir. We show that this leads to a dynamics that is a quantum dynamical semigroup and that has the correct final state.

\smallskip

Our paper is organized as follows. In Section \ref{mainressect} we give the setup of the problem, state our assumptions and present the main result, Theorem \ref{mainthm}. At the beginning of Section \ref{mainsect} we explain the mathematical description of the reservoir and, in Subsection \ref{subsconstruction} we construct the renormalized quantities (i.e., the system reference state). We provide the proof of Theorem \ref{mainthm} in Subsection \ref{subsdynamics} (representation of the dynamics by $\tau^t$) and Subsection \ref{cpsect} (complete positivity).

\section{Main result}
\label{mainressect}

The Hilbert space of a finite dimensional quantum system $\s$ in contact with a bosonic quantum field (reservoir) $\r$ is  
\begin{equation}
	\label{hilbspace}
{\mathcal H} = {\mathcal H}_\S\otimes{\mathcal H}_\R,
\end{equation}
where ${\mathcal H}_\s ={\mathbb C}^d$ and ${\mathcal H}_\r   = \oplus_{n\ge 0} L_{{\rm sym}}^2({\mathbb R}^{3n},d^{3n}k)$ 
is the Fock space over the single particle space $L^2({\mathbb R}^3,d^3k)$. We consider Hamiltonians 
\begin{equation}
	\label{hamilt}
H = H_\s +H_\r +\lambda\,  V_\s\otimes\varphi(g),
\end{equation}
where $H_\s$ and $V_\s$ are self-adjoint matrices on ${\mathcal H}_\s$, 
\begin{equation}
	\label{intro2}
H_\s = \sum_{j=1}^d E_j |\phi_j\rangle\langle\phi_j|,\quad \mbox{and} \quad H_\r = \int_{{\mathbb R}^3} |k|\, a^*(k)a(k) d^3k
\end{equation}
is the second quantization of multiplication with the function $|k|$, the energy of the mode $k$. The creation operators $a^*(k)$ and annihilation operator $a(k)$ satisfy the Bose canonical commutation relations $[a(k),a^*(l)]=\delta(k-l)$ (Dirac delta). We assume for convenience of exposition that all eigenvalues $E_j$ of $H_\s$ are simple -- our arguments are readily generalized to degenerate spectrum. The interaction strength is gauged by the coupling constant $\lambda\in\mathbb R$ and 
$$
\qquad \varphi(g) = \frac{1}{\sqrt 2}\big( a^*(g)+a(g)\big), \qquad a^*(g)=\int_{{\mathbb R}^3} g(k) a^*(k) d^3k,
$$
is the field and the creation operator (whose adjoint is $a(f)$, the annihilation operator), respectively, smoothed out with a form factor $g\in L^2({\mathbb R}^3,d^3k)$. 

In this work, we are concerned with the time evolution of observables $X\in{\mathcal B}({\mathcal H}_\s)$ under the coupled system-reservoir Heisenberg dynamics $\alpha_\lambda^t$ generated by the Hamiltonian $H$,
\begin{equation}
	\label{intro1}
t\mapsto \omega\big(\alpha^t_\lambda(X\otimes\bbbone_\r)\big).
\end{equation}
The initial state $\omega$ is a ``normal state", characterized by the fact that asymptotically in space, the reservoir is in its thermal equilibrium state. We do not demand that the system and reservoir are initially disentangled. There is a slight mathematical complication in the precise definition of \eqref{intro1} because thermal reservoirs are spatially infinitely extended systems. We explain this point in Section \ref{mathcsect}.

The non-interacting dynamics is the product $\alpha^t_0 = \alpha^t_\S\otimes\alpha^t_\R$, where the individual dynamics of each factor is generated by its own Hamiltonian $H_\s$ or $H_\r$. For small coupling constants $\lambda$ one can use a perturbation theory for the reduced system dynamics. Effectively, the energy levels of $H_\s$ acquire {\em complex valued} corrections (of $O(\lambda^2)$) which describe irreversibility of the open system dynamics. It is convenient to express this scheme in terms of the system {\em Liouville operator}
\begin{equation}
	\label{ls}
L_\s= H_\s\otimes\bbbone_\s - \bbbone_\s\otimes H_\s
\end{equation}
acting on the doubled space ${\mathcal H}_\s\otimes{\mathcal H}_\s$. The Liouville representation is quite standard \cite{Muka}. The eigenvalues of $L_\s$ are the differences of those of $H_\s$. They describe the temporal oscillations of the system density matrix elements in the energy basis under the uncoupled dynamics. Namely, the density matrix elements oscillate in time with frequencies that are the eigenvalues of $L_\s$. The coupling with the reservoir produces corrections. To lowest order in $\lambda$, the corrected eigenvalues are those of  $L_\s+\lambda^2\Lambda$, where $\Lambda$ is the so-called {\em level shift operator}, a non-selfadjoint matrix on ${\mathcal H}_\s\otimes{\mathcal H}_\s$, which can be calculated explicitly (c.f. \eqref{originallso}). The operators $L_\s$ and $\Lambda$ commute and satisfy
\begin{equation}
	\label{intro3}
(L_\s+\lambda^2\Lambda)\Omega_{\s,\beta}=0
\end{equation}
for all $\lambda\in{\mathbb R}$, where (c.f. \eqref{intro2})
\begin{equation}
	\label{gibbsvect}
\Omega_{\s,\beta} = Z_\s^{-1/2} \ \sum_j \e^{-\beta E_j/2}\phi_j\otimes\phi_j, \qquad Z_\s={\rm Tr}\, \e^{-\beta H_\s} 
\end{equation}
is the system Gibbs (equilibrium) vector, defining the equilibrium state
\begin{equation}
	\label{gibbsequilstate}
\omega_{\s,\beta}(X) = Z^{-1}_\s{\rm Tr}_\s(\e^{-\beta H_\s}X) = \SPn{\Omega_{\s,\beta}}{(X\otimes\bbbone_\s) \Omega_{\s,\beta}},\qquad X\in{\mathcal B}({\mathcal H}_\s).
\end{equation}
The relation \eqref{intro3} reflects the fact that the system Gibbs state is invariant under the coupled dynamics, to lowest order in the perturbation. (In fact, generically, it is the final system state, as $t\rightarrow\infty$, to lowest order in $\lambda$.) For simplicity of exposition, we assume that
\begin{itemize}
	\item[{\bf (A1)}] 	
	(i) \,\,All eigenvalues of $\Lambda$ are simple.\\
	(ii) All eigenvalues of $\Lambda$ but zero have strictly positive imaginary part,
	\begin{equation}
		\label{intro4}
		\gamma = \min\big\{ {\rm Im}\, a\ :\ a\in {\rm spec}(\Lambda)\backslash\{0\}  \big\} >0. 
	\end{equation}
\end{itemize}
 Since $L_\s$ and $\Lambda$ commute, the eigenvalues of $L_\s+\lambda^2\Lambda$ are of the form $e+\lambda^2 a$, with $e\in{\rm spec}(L_\s)$, $a\in{\rm spec}(\Lambda)$. In particular, for small enough, but non-vanishing $\lambda$, the operator $L_\s+\lambda^2\Lambda$ has only simple eigenvalues and, apart from zero, all its spectrum has imaginary part $\ge \gamma$. Both assumptions are readily and generically verified in concrete examples. Assumption (i) simplifies the analysis somewhat and guarantees in particular that $L_\s+\lambda^2\Lambda$ is diagonalizable. Assumption (ii) is commonly referred to as the {\em Fermi Golden Rule Condition} and ensures that irreversible effects are visible already in the lowest order correction to the dynamics.

 \medskip

In the dynamical theory of quantum resonances, the resonances (complex energy eigenvalues) associated to the Liouville operator are determined using spectral deformation- or Mourre theory \cite{JP0,JP1,BFS,Metal,Mappli}. In order not to muddle the core ideas of the current work, we follow here the technically least complicated situation, where the Hamiltonian is ``translation deformation analytic" \cite{JP0,BFS}. This requires a regularity assumption on the form factor $g\in L^2({\mathbb R}^3,d^3k)$. To state it, define the complex valued function $g_\beta$ on ${\mathbb R}\times S^2$ by
\begin{equation}
	g_\beta(u,\Sigma) = \sqrt{\frac{u}{1-e^{-\beta u}}}\ |u|^{1/2} \left\{
	\begin{array}{ll}
		g(u,\Sigma), & u\geq 0\\
		-\overline{g}(-u,\Sigma), & u<0
	\end{array}
	\right.
	\label{A2}
\end{equation}
where $g(v,\Sigma)$ is $g(k)$ in spherical coordinates $(v,\Sigma)\in{\mathbb R}_+\times S^2$. The regularity condition is the following.
\begin{itemize}
	\item[{\bf (A2)}] For $\theta\in\mathbb R$, define $(T_\theta g_\beta)(u,\Sigma) = g_\beta(u-\theta,\Sigma)$. There is a $\theta_0>0$ such that, viewed as a map from $\mathbb R$ to $L^2({\mathbb R}\times S^2,d u\times d\Sigma)$, the function $\theta\mapsto T_\theta g_\beta$ has an analytic extension to $0<{\rm Im}\theta <2\theta_0$ which is continuous as ${\rm Im}\theta\rightarrow 0_+$.
\end{itemize}
This condition is satisfied for instance for the following family of form factors, given in spherical coordinates  $(r,\Sigma)\in {\mathbb R}_+\times S^2={\mathbb R}^3$,
$$
g(k)=g(r,\Sigma) = r^p\e^{-r^m}g_1(\Sigma), 
$$
where $p=-1/2+n$, $n=0,1,2,\ldots$, $m=1,2$ and $g_1(\sigma)=\e^{\i\phi}\bar g_1(\sigma)$ for an arbitrary phase $\phi$ (see also \cite{FM}).

\medskip

Let $\alpha^t_\lambda$ be the coupled system-reservoir dynamics. The resonance approach gives the following expansion if $0<|\lambda|<\lambda_0$ for a sufficiently small $\lambda_0$. For all  system-reservoir initial states $\omega_0$ belonging to a dense set ${\mathcal S}_0$, all system observables $X\in{\mathcal B}({\mathcal H}_\S)$ and all times $t\ge 0$, 
\begin{equation}
	\label{01.1}
	\omega_0\big( \alpha^t_\lambda(X\otimes \bbbone_\R)\big) = \omega_{\S\R,\beta,\lambda}(X\otimes \bbbone_\R) +\omega_0\big( \delta^t_\lambda(X)\otimes\bbbone_\R\big)+R_{\lambda,t}(X),
\end{equation}
where $\omega_{\S\R,\beta,\lambda}$ is the coupled system-reservoir equilibrium state, where $\delta^t_\lambda: {\mathcal B}({\mathcal H}_\S)\rightarrow {\mathcal B}({\mathcal H}_\S)$ and where the remainder $R_{\lambda,t}(X)$ satisfy
\begin{eqnarray}
|\delta^t_\lambda(X)| &\le& C\e^{-\lambda^2\gamma t} \, \|X\|\\
\label{02.0}
| R_{\lambda,t}(X) |&\le& C\, |\lambda|\, \big( \e^{-\theta_0 t} + \lambda^2 t \, \e^{-\lambda^2(1+O(\lambda)) \gamma t}\big)\,\|X\|. 
\label{02.1}
\end{eqnarray}
Here, $\gamma$ is the gap \eqref{intro4} and $\theta_0$ is given in Assumption (A2).\footnote{If the initial state is of product form $\omega_\s\otimes\omega_{\R,\beta}$, then the term $ C |\lambda|\e^{-\theta_0 t}$ in \eqref{02.1} can be replaced by $ C \lambda^2\e^{-\theta_0 t}$, see Theorem 3.1 of \cite{Metal}, {\em Resonance theory of decoherence and thermalization}.} The resonance approach requires that $\lambda^2\gamma <\!\!< \theta_0$. The map $\delta^t_\lambda$ is defined by the relation
\begin{equation}
\label{05}
\big( \delta^t_\lambda(X)\otimes\bbbone_\S \big) \Omega_{\S,\beta} = \e^{\i t (L_\S+\lambda^2\Lambda)}P^\perp_{\S,\beta}\,  \big(X\otimes\bbbone_\S\big) \Omega_{\S,\beta},
\end{equation}
where $\Omega_{\S,\beta}\in {\mathcal H}_\S\otimes{\mathcal H}_\S$ is the system Gibbs vector \eqref{gibbsvect},  $P_{\S,\beta}=|\Omega_{\S,\beta}\rangle\langle \Omega_{\S,\beta}|$ and $P_{\S,\beta}^\perp =\bbbone_\S-P_{\S,\beta}$.\footnote{$\Omega_{\S,\beta}$ is {\em cyclic}, meaning that $({\mathcal B}({\mathcal H}_\S)\otimes\bbbone_\S)\Omega_{\S,\beta}={\mathcal H}_\S\otimes{\mathcal H}_\S$ and $\Omega_{\S,\beta}$ is {\em separating}, meaning that if $(X\otimes\bbbone_\S)\Omega_{\S,\beta}=0$ then $X=0$. Due to the cyclic and separating property, \eqref{05} defines the map $\delta^t_\lambda$ uniquely, and it shows that $t\mapsto\delta^t_\lambda$ is a group.} The operators $L_\S$ and $\Lambda$ are the system Liouville- and the level shift operators, respectively,  acting on ${\mathcal H}_\S\otimes{\mathcal H}_\S$ and commuting with each other. Under typical well-coupledness conditions (e.g. ``the Fermi Golden Rule condition") one has
\begin{equation}
	\label{08.1}
{\rm Ker}(L_\S+\lambda^2\Lambda) = {\mathbb C}\Omega_{\S,\beta},
\end{equation}
which sharpens \eqref{intro3}. The property of return to equilibrium follows from \eqref{01.1}, namely, 
\begin{equation}
\label{06}
\lim_{t\rightarrow \infty} \omega_0(\alpha^t_\lambda(X\otimes\bbbone_\R)) = \omega_{\S\R,\beta,\lambda}(X\otimes\bbbone_\R).
\end{equation}
By modifying $\delta^t_\lambda$, \eqref{05}, on the ``stationary subspace" ${\rm Ran}\, P_{\S,\beta}$, we define the map $\sigma^t_\lambda: {\mathcal B}({\mathcal H}_\S)\rightarrow  {\mathcal B}({\mathcal H}_\S)$  by 
\begin{equation}
\label{08}
\big( \sigma^t_\lambda(X)\otimes\bbbone_\S \big) \Omega_{\S,\beta} = \e^{\i t (L_\S+\lambda^2\Lambda)}\,  \big(X\otimes\bbbone_\S\big) \Omega_{\S,\beta}.
\end{equation}
It follows from \eqref{05} and \eqref{intro3} that
\begin{equation}
	\label{08.01}
	\sigma_t^\lambda(X) = \delta_t^\lambda (X) +\omega_{\s,\beta}(X)\bbbone_\s,\qquad X\in {\mathcal B}({\mathcal H}_\s).
\end{equation}
Expanding the joint equilibrium state for small $\lambda$, 
\begin{equation}
	\label{07}
	\omega_{\S\R,\beta,\lambda}(X\otimes \bbbone_\R) = \omega_{\S,\beta}(X) +R'_\lambda(X),
\end{equation}
where
\begin{eqnarray}
	| R'_\lambda(X)| \le C\lambda^2 \|X\|,
	\label{02.3}
\end{eqnarray} 
and combining \eqref{01.1} and \eqref{08.01} we obtain
\begin{equation}
	\label{01}
	\omega_0\big( \alpha^t_\lambda(X\otimes \bbbone_\R)\big) = \omega_0\big( \sigma^t_\lambda(X)\otimes\bbbone_\R\big) +R_\lambda'(X)+ R_{\lambda,t}(X),
\end{equation}
where $R_{\lambda,t}(X)$ and  $R'_\lambda(X)$ satisfy \eqref{02.1} and \eqref{02.3}, respectively.

 The expansion \eqref{01} has an advantage and a disadvantage over the expansion \eqref{01.1}. 
\begin{itemize}
	\item The advantage. The main term in \eqref{01} is given by $\sigma^t_\lambda$, which is a {\em quantum dynamical semigroup} (in the Heisenberg picture). This means that $\sigma^t_\lambda$ is a semigroup of completely positive maps and satisfies $\sigma^t_\lambda(\bbbone_\S)=\bbbone_\S$. The latter property (which is equivalent to the dual map acting on density matrices being trace preserving) follows directly from the definition \eqref{08} and \eqref{08.1}. We give a derivation of its complete positivity is Section \ref{cpsect}.

	\item The disadvantage. The main term of expansion \eqref{01} describes the time-asymptotics only up to an accuracy of $O(\lambda^2)$. Indeed, the final state of $\sigma^t_\lambda$ is the uncoupled equilibrium $\omega_{\S,\beta}$ while the true final state is the reduction to $\S$ of the coupled state $\omega_{\S\R,\beta,\lambda}$, as correctly described by \eqref{01.1}. In other words, the remainder in expansion \eqref{01} does not vanish as $t\rightarrow\infty$, but stays of $O(\lambda^2)$.
\end{itemize}

The main result of this paper is Theorem \ref{mainthm} below, which gives an effective system dynamics $\tau^t_\lambda$ that {\em combines the advantages} of the above expansions \eqref{01.1} and \eqref{01}, namely, 
\begin{itemize}
	\item[(i)] $\tau^t_\lambda$ is a quantum dynamical semigroup of the system, and 
	\item[(ii)] $\tau^t_\lambda$ describes the correct long time asymptotics (vanishing remainder as $t\rightarrow\infty$).
\end{itemize}

\begin{thm}
\label{mainthm}
There is a constant $\lambda_0>0$ such that the following holds for $|\lambda|<\lambda_0$. 
There is a completely positive, identity preserving semigroup $\tau^t_\lambda$ acting on ${\mathcal B}({\mathcal H}_\s)$, the observables of the system, such that $\forall \omega_0\in{\mathcal S}_0$,  $\forall t\ge 0$, $\forall X\in{\mathcal B}({\mathcal H}_\s)$, 
\begin{equation}
	\label{011}
	\omega_0\big(\alpha^t_\lambda (X\otimes\bbbone_\R)\big) =\omega_0\big(\tau^t_\lambda(X)\otimes\bbbone_\R\big) +R_{\lambda,t}(X),
\end{equation}
where $R_{\lambda,t}(X)$ satisfies 
\begin{equation}
		\label{finalremainder}	
		|R_{\lambda,t}(X)| \le C|\lambda|\, \big(1+\lambda^2 t\big)\,\e^{-\lambda^2(1+O(\lambda))\gamma t}\,   \|X\|.
\end{equation}	
The dynamical semigroup $\tau^t_\lambda$ can be constructed perturbatively in $\lambda$, by using the resonance data (energies and vectors) of a renormalized, $\lambda$-dependent system Hamiltonian.
\end{thm}

In analogy with \eqref{05} and \eqref{08}, we will construct $\tau^t_\lambda$ by the definition
\begin{equation}
	\label{012}
	\big(\tau^t_\lambda(X)\otimes\bbbone_\S\big) \widetilde\Omega_{\S,\beta,\lambda} = \e^{\i t(\widetilde L_\S +\lambda^2\widetilde \Lambda)} (X\otimes\bbbone_\S)\widetilde\Omega_{\S,\beta,\lambda},
\end{equation}
where $\widetilde L_\S=\widetilde L_\S(\lambda)$ and $\widetilde \Lambda=\widetilde\Lambda(\lambda)$ are suitably renormalized Liouville- and level shift operators, respectively, which commute with each other. Here, $\widetilde\Omega_{\S,\beta,\lambda}$ is a cyclic and separating vector spanning the kernel of $\widetilde L_\S +\lambda^2\widetilde \Lambda$.

\section{The renormalized dynamics and complete positivity.} 
\label{mainsect}

\subsection{States and dynamics}
\label{mathcsect}

The description \eqref{hilbspace}-\eqref{intro2} given above is common in the theoretical physics literature and serves, in particular, to introduce the system-reservoir interaction operators (taken here to be linear in the field, \eqref{hamilt}). It is, however, well known that the Fock space ${\mathcal H}_\r$ above is not the correct Hilbert space on which one can represent the state of a spatially infinitely extended bose gas in thermal equilibrium. To find that Hilbert space, one has to first perform the thermodynamic limit of the reservoir equilibrium state and then reconstruct its Hilbert space representation using the Gelfand-Naimark-Segal construction. This is the Araki-Woods representation for thermal reservoirs \cite{AW}.

It consists of a triple $({\mathcal H}_{\r,\beta}, \pi_\beta,\Omega_{\r})$, where ${\mathcal H}_{\r,\beta}$ is the representation Hilbert space, $\pi_\beta: {\mathfrak W}\rightarrow {\mathcal B}({\mathcal H}_{\r,\beta})$ is a representation of the Weyl algebra and $\Omega_\r\in{\mathcal H}_{\r,\beta}$ is a normalized vector representing the equilibrium state. Explicitly,
\begin{equation}
{\mathcal H}_{\r,\beta} = {\mathcal F}\big(L^2({\mathbb R}\times S^2,du\times d\Sigma)\big)\equiv \bigoplus_{n\ge 0} L^2_{\rm sym}\big(  ({\mathbb R}\times S^2)^n,(du\times d\Sigma)^n \big)
\end{equation}
is the bosonic Fock space over the single particle space $L^2({\mathbb R}\times S^2,du\times d\Sigma)$, where $d\Sigma$ is the uniform measure on the sphere $S^2$. The vector $\Omega_\r$ is the vacuum vector of $\mathcal F$ and the representation map is given by 
\begin{equation}
	\label{052}
\pi_\beta(W(f)) =W(f_\beta),
\end{equation}
where $f\mapsto f_\beta$ was defined by \eqref{A2}. The operator $W(f)$ on the left side of \eqref{052} is an (abstract) Weyl operator in $\mathfrak W$, while the represented $W(f_\beta)$ on the right side is given by $W(f_\beta)=\e^{\i\varphi(f_\beta)}$, with $\varphi(f_\beta) =2^{-1/2}[a(f_\beta)+a^*(f_\beta)]$. Here, $a^*(f_\beta)$ is the creation operator smoothed out with $f_\beta$, acting on $\mathcal F$ and $a(f_\beta)$ its adjoint. The reservoir equilibrium state at temperature $T=1/\beta$ is represented as
$$
\omega_{\r,\beta}(W(f)) =\SPn{\Omega_\r}{\pi_\beta(W(f))\Omega_\r}.
$$
The free reservoir dynamics is implemented as $\pi_\beta(W(\e^{\i\omega t}f)) = \e^{\i tL_\r}\pi_\beta(W(f))\e^{-\i t L_\r}$, where
\begin{equation}
	\label{lr}
	L_\r = d\Gamma(u)
\end{equation}
is the reservoir Liouville operator, the second quantization of multiplication with the radial variable $u$. 

Together with \eqref{ls}, the joint system-reservoir Hilbert space and non-interacting Liouville operator are given by 
\begin{equation}
\label{sr}
{\mathcal H} = {\mathbb C}^d\otimes{\mathbb C}^d\otimes {\mathcal H}_{\r,\beta}\qquad \mbox{and}\qquad L_0=L_\s+L_\r.
\end{equation}
The interaction associated with \eqref{hamilt} is represented by the operator
\begin{equation}
	\label{intop}
	I = V_\s\otimes\bbbone_\s\otimes \varphi(g_\beta) -J\big(V_\s\otimes\bbbone_\s\otimes \varphi(g_\beta)  \big)J,
\end{equation}
where $J=J_\s\otimes J_\r$ is the modular conjugation. It is given explicitly as follows. Let ${\mathcal C}$ be the anti-linear operator acting on ${\mathbb C}^d$ by taking complex conjugates of vector coordinates in the energy basis $\{\varphi_n\}$, then $J_\s$ acts on ${\mathbb C}^d\otimes {\mathbb C}^d$ as $J_\s \chi\otimes\psi = {\mathcal C}\psi\otimes{\mathcal C}\chi$. Similarly, $J_\r$ acts on $\mathcal F$ sector-wise and on the $n$-sector, its action is $J_\r\psi_n(u_1,\Sigma_1,\ldots,u_n,\Sigma_n) = \overline{\psi_n(-u_1,\Sigma_1,\ldots,-u_n,\Sigma_n)}$. The full Liouville operator is then
\begin{equation}
\label{liouvilleop}
L_\lambda = L_0+\lambda I.
\end{equation}
The non-interacting and interacting systems, whose dynamics is generated by $L_0$ and $L_\lambda$, have unique $\beta$-KMS states $\omega_{\s\r,\beta,0}$ and $\omega_{\s\r,\beta,\lambda}$, which are represented by the KMS vectors $\Omega_{\s\r,\beta,0}$ and $\Omega_{\s\r,\beta,\lambda}$ respectively, where (recall \eqref{gibbsvect})
\begin{equation}
	\label{twoKMS}
\Omega_{\s\r,\beta,0} = \Omega_{\s,\beta}\otimes\Omega_\r\qquad \mbox{and}\qquad \Omega_{\s\r,\beta,\lambda} = \frac{\e^{-\beta(L_0+\lambda V_\s\otimes\bbbone_\s\otimes\varphi(g_\beta))/2}\Omega_{\s\r,\beta,0}}{\| \e^{-\beta(L_0+\lambda V_\s\otimes\bbbone_\s\otimes\varphi(g_\beta))/2}\Omega_{\s\r,\beta,0}  \|}. 
\end{equation}
We refer to \cite{BR,DJP,BFS} for more detail on the construction of the interacting KMS state.

\subsection{Construction of the renormalized quantities}
\label{subsconstruction}

The reduction of the joint equilibrium state to the system is given by the density matrix $\rho_{\S,\beta,\lambda}$, defined by
\begin{equation}
\label{013}
{\rm Tr}_\S\big( \rho_{\S,\beta,\lambda}X\big) =\omega_{\S\R,\beta,\lambda}(X\otimes\bbbone_\R)\qquad\mbox{for all $X\in{\mathcal B}({\mathcal H}_\S)$.}
\end{equation}
Since $\omega_{\S\R,\beta,\lambda}$ is faithful, it is readily seen that $\rho_{\S,\beta,\lambda}$ is {\em strictly} positive. Set
\begin{equation}
\label{zdef}
\widetilde Z :=\|\rho_{\S,\beta,\lambda}\|^{-1}
\end{equation}
(operator norm) and define the renormalized system Hamiltonian by
\begin{equation}
	\label{014.1}
	\widetilde H_\S = -\tfrac1\beta \ln\big(\widetilde Z \rho_{\S,\beta,\lambda} \big),
\end{equation}
so that 
\begin{equation}
\rho_{\S,\beta,\lambda} = \widetilde Z^{-1}\,  \e^{-\beta \widetilde H_\S}.
\end{equation}
Note that $\widetilde Z={\rm Tr}_\s\,\e^{-\beta \widetilde H_\s}$. The operator $\widetilde H_\S$ depends on $\lambda$ and we have $\widetilde H_\S|_{\lambda=0} =H_\S$ and $\widetilde Z|_{\lambda=0} = {\rm Tr}_\s\,\e^{-\beta H_\s}\equiv Z_\s$ (c.f. \eqref{gibbsvect}). 

\begin{lem}
	\label{lem001}
Let $\{\phi_n\}_{n=1,\ldots,d}$ be an orthonormal basis of eigenvectors of $H_\S$, such that $H_\S\phi_n=E_n\phi_n$. The eigenvalues of $\widetilde H_\S$ are $\widetilde E_n$, satisfying $E_n-\widetilde E_n=O(\lambda)$. The normalized eigenvectors, $\wt H_\s\wt\phi_n=\wt E_n\wt\phi_n$, satisfy $\phi_n-\wt\phi_n =O(\lambda)$.  
\end{lem}

{\em Proof of Lemma \ref{lem001}.\ } Araki's perturbation theory of KMS states \cite{BR,DJP,BFS} yields $\|\rho_{\s,\beta,\lambda} -\rho_{\s,\beta,0}\| = O(\lambda)$. 
It follows from \eqref{zdef} that $|\widetilde Z- Z_\s|=O(\lambda)$, where $Z_\s$ is the unperturbed system partition function, \eqref{gibbsvect}. Then \eqref{014.1} gives $\|\widetilde H_\s-H_\s\|=O(\lambda)$. The lemma then follows from usual analytic perturbation theory for matrices.\hfill\qed

We define
\begin{eqnarray}
\widetilde L_\S &=& \widetilde H_\S\otimes\bbbone_\S - \bbbone_\S\otimes {\mathcal C}\widetilde  H_\S{\mathcal C} \label{014}\\
\widetilde\Omega_{\S,\beta,\lambda} &=& \widetilde Z^{-1/2} \sum_{n=1}^d \e^{-\beta\widetilde E_n/2} \widetilde\phi_n\otimes{\mathcal C}\widetilde\phi_n,
\end{eqnarray}
where $\mathcal C$ is the antilinear map satisfying ${\mathcal C}\phi_n=\phi_n$ (i.e., $\mathcal C$ implements complex conjugation of coordinates in the basis $\{\phi_n\}$). The vector $\widetilde\Omega_{\S,\beta,\lambda}$ represents the state $\rho_{\S,\beta,\lambda}$, meaning
\begin{equation}
	\label{repstate}
	\langle\wt\Omega_{\s,\beta,\lambda}, (X\otimes\bbbone_\s)\wt\Omega_{\s,\beta,\lambda} \rangle = {\rm Tr}_\s\, (\rho_{\s,\beta,\lambda} X),\qquad X\in{\mathcal B}({\mathcal H}_\s).
\end{equation}
$\Omega_{\s,\beta,\lambda}$ and is a $\beta$-KMS vector with respect to the dynamics $\e^{\i t\widetilde L_\S} \cdot \e^{-\i t\widetilde L_\S}$ of the system observable algebra  ${\mathcal B}({\mathcal H}_\S)\otimes\bbbone_\S$.  We let 
\begin{eqnarray}
\label{015}
\widetilde L_0 &=& \widetilde L_\S +L_\R\\
\widetilde\Omega_0 &=& \widetilde\Omega_{\S,\beta,\lambda}\otimes\Omega_\R
\label{0015}
\end{eqnarray}
and denote by $\widetilde P_{\widetilde e}$ the eigenprojection onto the eigenvalue $\widetilde e$ of $\widetilde L_0$.

The level shift operators of the original (not renormalized) system are given as follows. For each $e\in{\rm spec}(L_\S)$, 
\begin{equation}
	\label{originallso}
\Lambda_e = -P_e I P^\perp_e (L_0-e+\i0_+)^{-1} I P_e,\qquad \mbox{and}\qquad \Lambda = \sum_{e\in{\rm spec}(L_\S)}\Lambda_e,
\end{equation}
where $P_e$ is the spectral projection of $L_0$ onto the eigenvalue $e$ (having multiplicity $m_e$). $\Lambda_e$ is diagonalizable and has the spectral representation
\begin{equation}
	\label{048}
\Lambda_e =\sum_{j=1}^{m_e} \lambda_{e,j} Q_{e,j},
\end{equation}
where $Q_{e,j}=P_e Q_{e,j}=Q_{e,j}P_e$ and $\lambda_{e,j}$ are the spectral projections and eigenvalues, which are all simple (Assumption (A1)(i)). According to Assumption (A1)(ii), we have ${\rm ker}\Lambda = {\mathbb C}\Omega_{\S,\beta}$ and ${\rm Im }\lambda_{e,j} >0$ for all $e\neq 0$ and $j$, ${\rm Im }\lambda_{0,j} >0$ for $j=1,\ldots,d-1$ and $\lambda_{0,d}=0$ (one-dimensional kernel of $\Lambda$). 

We now define the level shift operator $\wt\Lambda$ of the renormalized system. For each $\widetilde e\in{\rm spec}(\widetilde L_\s)$, set
\begin{equation}
\label{016}
\widetilde\Lambda_{\widetilde e} =- \widetilde P_{\widetilde e} I \wt P_{\widetilde e}^\perp (\widetilde L_0-\widetilde e+\i 0_+)^{-1} I\wt P_{\widetilde e}\, ,
\qquad \mbox{and}\qquad \widetilde\Lambda = \sum_{\widetilde e \in {\rm spec}(\widetilde L_\S)} \widetilde\Lambda_{\widetilde e}\widetilde P_{\widetilde e}\, .
\end{equation}

\begin{prop}
	\label{prop2}
	The operator $\wt\Lambda_{\wt e}$ exists for each $\wt e\in{\rm spec}(\wt L_\S)$ and satisfies $\Lambda_e-\wt\Lambda_{\wt e}=O(\lambda)$. Its spectrum consists of simple eigenvalues $\wt\lambda_{\wt e,j}$, $j=1,\ldots,m_e$, satisfying $\lambda_{e,j}-\wt\lambda_{\wt e,j} =O(\lambda)$. The associated Riesz spectral projections $\wt Q_{\wt e,j}$ satisfy $Q_{e,j}-\wt Q_{\wt e,j}=O(\lambda)$. Moreover, ${\rm ker}\widetilde \Lambda={\mathbb C} \widetilde \Omega_0$.
\end{prop}
The proposition implies that $\wt\Lambda$ has the spectral representation
 \begin{equation}
 	\label{022}
 	\wt\Lambda=  \sum_{j=1}^{d-1} \wt{\lambda}_{0,j}\, \wt{Q}_{0,j} +\sum_{\widetilde e\not=0} \sum_{j=1}^{m_e}  \wt{\lambda}_{\wt{e},j}\, \wt{Q}_{\wt{e},j}.
 \end{equation}

{\em Proof of Proposition \ref{prop2}.\ } Let $U_\theta=\e^{\i\theta d\Gamma(-\i\partial_u)}$, so that $U_\theta L_0 U_\theta^* = L_0+\theta N$, where $N=d\Gamma(\bbbone_\R)$ is the number operator. Setting $I_\theta=U_\theta I U_\theta^*$ and using that $U_\theta P_e=P_eU_\theta=P_e$, we have for all $\theta\in\mathbb R$ and $\epsilon>0$
\begin{eqnarray*}
P_eI(L_0-e+\i\epsilon)^{-1} IP_e&=& P_e I_\theta (L_0+\theta N-e+\i \epsilon)^{-1} I_\theta P_e.
\end{eqnarray*}
By assumption (A2), the right side has an analytic extension into values of $\theta$ in a strip with ${\rm Im}\,\theta<2\theta_0$, for some $\theta_0>0$ and so
\begin{eqnarray*}
	P_eI(L_0-e+\i\epsilon)^{-1} IP_e&=& P_e I_{\i \theta_0} (L_0+\i\theta_0 N-e+\i \epsilon)^{-1} I_{\i\theta_0} P_e\\
	&=& \wt P_{\wt e} I_{\i \theta_0} (\wt L_0+\i\theta_0 N-\wt e+\i \epsilon)^{-1} I_{\i\theta_0} \wt P_{\wt e} +O(\lambda),
\end{eqnarray*}
where the error term bounded uniformly in $\epsilon>0$. As  $U_\theta \wt P_{\wt e}=\wt P_{\wt e}U_\theta=\wt P_{\wt e}$, we can undo the spectral deformation in the main term on the right side and take $\epsilon\rightarrow 0_+$ to obtain
\begin{equation}
\label{023}
\Lambda_e=\wt\Lambda_{\wt e}+O(\lambda). 
\end{equation}
The statements about the eigenvalues and Riesz eigenprojections follow from basic perturbation theory. (Recall that $\Lambda_e$ has simple, $\lambda$-independent eigenvalues by Assumption (A1).) To show that ${\rm ker}\wt\Lambda={\mathbb C}\wt\Omega_0$ it suffices to show that $\wt\Lambda_0\wt \Omega_0=0$, as all the eigenvalues $\wt\lambda_{\wt e,j}$ associated to $e\neq 0$ and for $e=0$ and $j=1,\ldots,d-1$, have strictly positive imaginary part, a property which is inherited from the eigenvalues of $\Lambda$ (for $\lambda$ small).

To show $\wt\Lambda_0\wt \Omega_0=0$ we introduce the auxiliary Liouville operator
\begin{equation}
\label{auxliouv}
	\wt L_\mu=\wt L_0+\lambda \mu  I,
\end{equation}
where $I$ is given in \eqref{intop}. By Araki's perturbation theory of KMS states, we know that
\begin{equation}
	\label{025}
\wt L_\mu \wt \Omega_\mu =0,
\end{equation}
where
\begin{equation}
\wt \Omega_\mu =\frac{\e^{-\beta\{\wt L_0 +\lambda\mu V_\s\otimes\bbbone_\s\otimes\varphi(g_\beta)\}/2} \wt\Omega_0}{\|\e^{-\beta\{\wt L_0 +\lambda\mu V_\s\otimes\bbbone_\s\otimes\varphi(g_\beta)\}/2} \wt\Omega_0\|}.
\end{equation}
\begin{lem}
	\label{lem3}
	Let $g_0>0$ be the spectral gap of $\wt L_\s$ at zero. The operator $\wt L_\mu^\perp:=\wt P_0^\perp \wt L_\mu\wt P_0^\perp|_{{\rm Ran}\wt P_0^\perp}$ has purely absolutely continuous spectrum in the open interval $(-g_0/2,g_0/2)$. In particular, zero is not an eigenvalue of $\wt L_\mu^\perp$. 
\end{lem}

{\em Proof of Lemma \ref{lem3}.\ } Let $\varphi$ be a $U_\theta$-analytic vector. For ${\rm Im} z<0$
\begin{eqnarray}
\scalprod{\varphi}{(\wt L_\mu^\perp -z)^{-1}\varphi} &=& \scalprod{\varphi_{\bar\theta}}{(\wt L_0^\perp +\theta N^\perp +\lambda\mu I_\theta^\perp  -z)^{-1}\varphi_\theta}\nonumber\\
&=&\scalprod{\varphi_{\bar\theta}}{(\wt L_0^\perp+\theta N^\perp-z)^{-1} \sum_{n\geq 0} (-\lambda\mu)^n \left[I_\theta^\perp(\wt L_0^\perp+\theta N^\perp-z)^{-1}\right]^n\varphi_\theta},
\label{032}
\end{eqnarray}
where $X^\perp = \wt P_0^\perp X\wt P_0^\perp|_{{\rm Ran}\wt P_0^\perp}$. Using the decomposition $\wt P^\perp_0 = \wt P^\perp_\S\otimes P_\R +\bbbone_\s\otimes P^\perp_\R$, where 
$\wt P_\s$ is the orthogonal projection onto the kernel of $\wt L_\s$
%$\wt P_\S= |\wt\Omega_{\S,\beta,\lambda}\rangle\langle\wt\Omega_{\S, \beta,\lambda}|$
 and $P_\R=|\Omega_\R\rangle\langle\Omega_\R|$, we easily obtain the bounds
\begin{eqnarray}
	\label{bounds1}
\| (\wt L_0^\perp+\i\theta_0 N^\perp-z)^{-1}\| &\le& \max\big\{ \max_{\wt e\neq 0} |\wt e-z|^{-1}, |\theta_0-{\rm Im} z|^{-1} \big\} \\
\|I_{\i \theta_0}^\perp(\wt L_0^\perp+\i\theta_0 N^\perp-z)^{-1}\|&\le& C_{\theta_0} \, \max\big\{ \max_{\wt e\neq 0} |\wt e-z|^{-1}, |\theta_0-{\rm Im} z|^{-1} \big\}.
\label{bounds2}
\end{eqnarray}
Thus, for ${\rm Im} z\le 0$ and $|{\rm Re} z|\le g_0/2$, where $g_0>0$ is the spectral gap of $\wt L_\S$ at zero, the combination of \eqref{032}, \eqref{bounds1} and \eqref{bounds2} gives the limiting absorption principle
$$
\sup_{z\, :\, |{\rm Re}z|\le g_0/2, {\rm Im}z\le 0}\left| \scalprod{\varphi}{(\wt L_\mu^\perp -z)^{-1}\varphi} \right| \le C(\varphi).
$$
This implies that $\wt L_\mu^\perp$ has purely absolutely continuous spectrum in the interval $(-g_0/2,g_0/2)$. Lemma \ref{lem3} is proven.

\medskip  

Combining \eqref{025} with Lemma \ref{lem3}, and invoking the isospectrality of the Feshbach map (see for instance Proposition B.2 in \cite{KM2}), we obtain
\begin{equation}
\label{026}
{\mathfrak F}(\wt L_\mu;\wt P_0) \wt P_0\wt\Omega_\mu =0,
\end{equation}
where
\begin{equation}
\label{027}
{\mathfrak F}(\wt L_\mu;\wt P_0) = -\lambda^2\mu^2\wt P_0 I \wt P_0^\perp (\wt L_\mu^\perp +\i 0_+)^{-1}I\wt P_0.
\end{equation}
We now use the translation analyticity to obtain
\begin{equation}
\label{030}
\wt P_0 I (\wt L_\mu^\perp +\i 0_+)^{-1}I\wt P_0= \wt P_0 I_{\i \theta_0} (\wt L_0^\perp +\i\theta_0N^\perp +\lambda\mu I^\perp_{\i\theta_0} )^{-1}I_{\i\theta_0}\wt P_0.
\end{equation}
Combining \eqref{026}, \eqref{027} and \eqref{030}, and taking $\mu\rightarrow 0$, gives
\begin{equation}
\label{031}
\wt P_0 I_{\i \theta_0} (\wt L_0^\perp +\i\theta_0 
N^\perp  )^{-1}I_{\i\theta_0}\wt P_0 \wt\Omega_0=0.
\end{equation}
Reversing the spectral deformation on the left hand side of \eqref{031} gives precisely $\wt\Lambda_0\wt\Omega_0=0$. This completes the proof of Proposition \ref{prop2}.\hfill \qed

\subsection{Representation of the dynamics: Proof of \eqref{011}}
\label{subsdynamics}

We first introduce the dense set of initial states for which the dynamical resonance theory based on spectral deformation can be applied. The three vectors $\wt\Omega_0$, $\Omega_{\s\r,\beta,0}$ and $\Omega_{\s\r,\beta,\lambda}$ play a role in what follows. We recall their definitions, \eqref{0015} and \eqref{twoKMS}. 

Let ${\mathfrak M}_0\subset\mathfrak M$ be the set of all finite linear combinations of operators of the form $\pi(A_\S\otimes W(f))$,  where $A_\S\in{\mathcal B}({\mathcal H}_\S)$ and $W(f)$ is a Weyl operator smoothed out with a test function $f$ that satisfies Assumption (A2). The following properties of the set of vectors $J{\mathfrak M}_0 \wt\Omega_0 = J{\mathfrak M}_0J \wt \Omega_0$  are not difficult to verify:
\begin{equation} 
\label{041}
\mbox{$J{\mathfrak M}_0 \wt\Omega_0$ is dense in $\mathcal H$\quad and}\quad  J{\mathfrak M}_0\wt\Omega_0\subset {\mathcal A}_{\theta_0}\cap{\rm Dom}(\e^{\alpha N}) \mbox{ for all $\alpha\in\mathbb R$}.
\end{equation}
Here, ${\mathcal A}_{\theta_0}$ is the set of vectors $\psi\in\mathcal H$ such that $\theta\mapsto \e^{\i \theta A}\psi$, $A=d\Gamma(-\i\partial_u)$, is analytic in $\theta$ in a strip ${\rm Im}\theta <\theta_0$ and  $N=d\Gamma(\bbbone_\r)$ is the number operator. We consider the set of states
\begin{equation}
\label{040}
{\mathcal S}_0 = \big\{ \omega(\cdot) = \langle{JC\wt\Omega_0},\pi(\cdot) JC\wt\Omega_0\rangle\ :\ C\in{\mathfrak M}_0  \big\}.
\end{equation}

\begin{lem}
\label{lem4}
We have $\wt\Omega_0=JD\Omega_{\s\r,\beta,\lambda}$ for an operator $D$ affiliated with ${\mathfrak M}$. Moreover, given any $\alpha>0$, we have for small enough $\lambda$, ${\rm Dom}(D^\#)\supset{\rm Dom}(\e^{\alpha N})$ and  $D^\#\e^{-\alpha N}{\mathcal A}_{\theta_0} \subset {\mathcal A}_{\theta_0}$. Here, $D^\#$ stands for $D$ or its adjoint $D^*$. Moreover, $\Omega_{\s\r,\beta,\lambda}\in {\mathcal A}_{\theta_0}$. 
\end{lem}
{\em Remark. } The vectors $\Omega_{\s\r,\beta,0}$, $\Omega_{\s\r,\beta,\lambda}$ and $\wt\Omega_0$   are all invariant under the action of the modular conjugation $J$. This implies that $\wt\Omega_0=D\Omega_{\s\r,\beta,\lambda}=JDJ\Omega_{\s\r,\beta,\lambda}$.

\medskip

{\em Proof of Lemma \ref{lem4}.\ }  
We first construct an operator $G$ satisfying  $\Omega_0=JG\Omega_{\s\r,\beta,\lambda}$ having the desired regularity  properties. 
The perturbation theory of KMS states gives 
\begin{equation}
	\label{043}
\Omega_{\s\r,\beta,\lambda} =c^{-1}  \e^{-\beta (L_0+\lambda K)/2} \e^{\beta L_0/2}\Omega_{\s\r,\beta,0} \quad\mbox{and}\quad \Omega_{\s\r,\beta,0}= c\e^{-\beta L_0/2} \e^{\beta(L_0+\lambda K)/2}\Omega_{\s\r,\beta,\lambda},
\end{equation} 
where $c$ is a normalization constant and, for short, $K=V_\s\otimes\bbbone_\s\otimes\varphi(g_\beta)$. A Dyson series expansion yields
\begin{equation}
	\label{042}
	 \e^{-\beta L_0/2}\e^{\beta (L_0+\lambda K)/2} =  \sum_{n\ge 0} \lambda^n \int_{0\le s_1\le\cdots\le s_n\le\beta/2} ds_1 \cdots ds_n \, K(s_n)\cdots K(s_1)=:c^{-1} G,
\end{equation}
where $K(s) = \e^{-s L_0} K \e^{sL_0}$. Using that 
$$
\sup_{0\le s\le\beta/2} \| \e^{\alpha N} K(s)\e^{-\alpha N} (N+1)^{-1/2}\| <\infty,
$$
one readily sees that, for any $\alpha>0$ fixed and $\lambda$ small enough, the series in \eqref{042} converges strongly on ${\rm Dom}(\e^{\alpha N})$ and defines an element affiliated with $\mathfrak M$, and that furthermore, ${\rm Ran}(G\e^{-\alpha N})\subset{\rm Dom}(\e^{\alpha N})$.  The analogous expansion and result can be obtained starting with $\e^{-\beta (L_0+\lambda K)/2} \e^{\beta L_0/2}$, which shows that $\Omega_{\s\r,\beta,\lambda}\in{\rm Dom}(\e^{\alpha N})$. Combining this with \eqref{043} gives
\begin{equation}
	\label{045}
\Omega_{\s\r,\beta,0}= G\Omega_{\s\r,\beta,\lambda} = JG\Omega_{\s\r,\beta,\lambda}.
\end{equation}
The last equality follows from $J\Omega_{\s\r,\beta,0}=\Omega_{\s\r\,\beta,0}$. The cyclicity of $\Omega_{\s,\beta}$ implies that there is a $D_\S\in{\mathcal B}({\mathcal H}_\s)$ satisfying  $\wt\Omega_0=J(D_\S\otimes\bbbone_\S\otimes\bbbone_\R)\Omega_{\s\r,\beta,0}$. Thus from \eqref{045},
\begin{equation}
\label{046}
\wt\Omega_0 = J (D_\S\otimes\bbbone_\S\otimes\bbbone_\R)G\Omega_{\s\r,\beta,\lambda}=:J D\Omega_{\s\r,\beta,\lambda}.
\end{equation}
It remains to prove the analyticity statement, which is the same as $G\e^{-\alpha N}{\mathcal A}_{\theta_0} \subset {\mathcal A}_{\theta_0}$. This follows again from the series expansion of $G$, \eqref{042}, and the fact that $\e^{\i\theta A} K(s_n)\cdots K(s_1) \e^{-\i\theta A} = K_\theta(s_n)\cdots K_\theta(s_1)$,
where $K_\theta(s) = \e^{\i \theta A} \e^{-s L_0} K\e^{s L_0}\e^{\i\theta A}$, is analytic.  

Finally, to show that $\Omega_{\s\r,\beta,\lambda}\in{\mathcal A}_{\theta_0}$, we note that the adjoint of the Dyson series expansion \eqref{042} gives that $G^*\e^{-\alpha N}{\mathcal A}_{\theta_0} \subset {\mathcal A}_{\theta_0}$ and the desired result follows from \eqref{043}. \hfill \qed

\bigskip
For $\omega_0\in{\mathcal S}_0$ and $X\in {\mathcal B}({\mathcal H}_\S)$, we have
\begin{eqnarray} \omega_0\big(\alpha^t_\lambda(X\otimes\bbbone_\R) \big) 
&=& \scalprod{JC\wt\Omega_0}{\e^{\i t L} (X\otimes\bbbone_\S\otimes\bbbone_\R)\e^{-\i tL}JC\wt\Omega_0}\nonumber\\
&=&\scalprod{JC^*C
	\wt\Omega_0}{\e^{\i tL}(X\otimes\bbbone_\S\otimes\bbbone_\R)\e^{-\i tL}\wt\Omega_0}\nonumber\\
&=& \scalprod{JC^*C
	\wt\Omega_0}{\e^{\i tL}(X\otimes\bbbone_\S\otimes\bbbone_\R)\e^{-\i tL} J DJ \Omega_{\s\r,\beta,\lambda}}\nonumber\\
&=&\scalprod{JC^*C
	\wt\Omega_0}{ JDJ \e^{\i tL}(X\otimes\bbbone_\S\otimes\bbbone_\R) \Omega_{\s\r,\beta,\lambda}}\nonumber\\
&=&\scalprod{JD^*C^*C
	\wt\Omega_0}{ \e^{\i tL}(X\otimes\bbbone_\S\otimes\bbbone_\R) \Omega_{\s\r,\beta,\lambda}}. 
\label{047}
\end{eqnarray}
Lemma \ref{lem4} gives $\Omega_{\s\r,\beta,\lambda}\in{\mathcal A}_{\theta_0}$ and since $C^*C\wt\Omega_0\in{\rm Dom}(\e^{\alpha N})\cap{\mathcal A}_{\theta_0}$, it also gives  $JD^*C^*C\wt\Omega_0\in{\mathcal A}_{\theta_0}$. Thus one can apply the spectral deformation method to \eqref{047} to obtain
\begin{align}
	\omega_0\big(\alpha^t_\lambda(X\otimes\bbbone_\R) \big) =& \scalprod{[JD^*C^*C
		\wt\Omega_0]_{\bar\theta}}{(|[\Omega_{\s\r,\beta,\lambda}]_{\theta}\rangle\langle[\Omega_{\s\r,\beta,\lambda}]_{\bar\theta}|) \, (X\otimes\bbbone_\S\otimes\bbbone_\R) [\Omega_{\s\r,\beta,\lambda}]_\theta}\nonumber\\
	&+ \sum_{j=1}^{d-1} \e^{\i t \lambda^2a_{0,j}} \scalprod{[JC^*C\wt\Omega_0]_{\bar\theta}}{\Pi_{0,j}(\theta) (X\otimes\bbbone_\S\otimes\bbbone_\R) [\Omega_{\s\r,\beta,\lambda}]_\theta}\nonumber\\
	&+\sum_{e\not=0} \sum_{j=1}^{m_e} \e^{\i t (e+\lambda^2a_{e,j})} \scalprod{[JC^*C\wt\Omega_0]_{\bar\theta}}{\Pi_{e,j}(\theta) (X\otimes\bbbone_\S\otimes\bbbone_\R) [\Omega_{\s\r,\beta,\lambda}]_\theta}+R(X,t),
	\label{017}
\end{align}
where $[\psi]_\theta=\e^{\i\theta A}\psi$ and (recall \eqref{048})
\begin{eqnarray}
\Pi_{e,j}(\theta) &=& Q_{e,j} +O(\lambda)\\
a_{e,j} &=& \lambda_{e,j} +O(\lambda).
\end{eqnarray}
The remainder $R(X,t)$ in \eqref{017} satisfies 
\begin{equation}
	\label{remest}
|R(X,t)| \le {\rm const.}  |\lambda| \big( \e^{-\theta_0 t} + \e^{-\lambda^2(1+O(\lambda))\gamma t} \big) \|X\| \le {\rm const.} |\lambda| \e^{-\lambda^2(1+O(\lambda))\gamma t}\, \|X\|.
\end{equation}
The contribution $\propto \e^{-\theta_0t}$ is the usual contour integral term (c.f. \cite{Metal}), the other term is due to the fact that in the summands decaying in time $t$, we can replace $D^*$ by $\bbbone$ plus a remainder of $O(\lambda)$ and we have $|\e^{\i t \lambda^2 a_{e,j}}|= \e^{-\lambda^2(1+O(\lambda))\gamma t}$. The first term on the right side of \eqref{017} equals
\begin{align}
	\lefteqn{
\scalprod{JD^*C^*C
	\wt\Omega_0}{\Omega_{\s\r,\beta,\lambda}} \scalprod{\Omega_{\s\r,\beta,\lambda}}{ (X\otimes\bbbone_\S\otimes\bbbone_\R) \Omega_{\s\r,\beta,\lambda}}}\qquad\qquad\qquad\qquad\qquad\qquad\qquad\qquad\nonumber\\ 
&= \scalprod{\Omega_{\s\r,\beta,\lambda}}{(X\otimes\bbbone_\S\otimes\bbbone_\R)\Omega_{\s\r,\beta,\lambda}}\nonumber\\
&= \scalprod{\wt\Omega_0}{(X\otimes\bbbone_\S\otimes\bbbone_\r)\wt\Omega_0}\nonumber\\
&=\scalprod{JC^*C\wt\Omega_0}{ (|\wt\Omega_0\rangle \langle\wt \Omega_0|)(X\otimes\bbbone_\S\otimes\bbbone_\R)\wt\Omega_0}.
\label{020}
\end{align}
In the first step, we have made use of 
$$
\langle JD^*C^*C\wt\Omega_0,\Omega_{\s\r,\beta,\lambda}\rangle = \langle J C^*C\wt\Omega_0, J D\Omega_{\s\r,\beta,\lambda}\rangle = \langle J C^*C\wt\Omega_0, \wt\Omega_0\rangle = \|C\wt\Omega_0\|^2=1
$$
and in the last step, again, $\langle JC^*C\wt\Omega_0, \wt\Omega_0\rangle=1$. The second equality in \eqref{020} follows from \eqref{013}, \eqref{repstate} and \eqref{0015}.

Since all ``directions"  but the stationary one in \eqref{017} are decaying, i.e., ${\rm Im}\, a_{e,j}>0$ for all $e,j$, we can replace in these terms in the exponents $e$ and $a_{e,j}$ by $\wt e$ and  $\widetilde\lambda_{\widetilde e,j}$, $\Pi_{e,j}$ by $\widetilde Q_{\widetilde e,j}$ and $\Omega_{\s\r,\beta,\lambda}$ by $\wt\Omega_0$ (see Proposition \ref{prop2}). This changes the remainder into a new one which, instead of \eqref{remest}, has the bound \eqref{finalremainder}.\footnote{Use the estimate $|\e^{-\lambda^2 t{\rm Im }a_{e,j}} - \e^{-\lambda^2 t{\rm Im }\widetilde\lambda_{e,j}}| = \e^{-\lambda^2 t{\rm Im }a_{e,j}}|1-\e^{\lambda^2 t{\rm Im }(a_{e,j}-\widetilde\lambda_{\widetilde e,j})}| \le {\rm const.} |\lambda|^3 t \e^{-\lambda^2(1+O(\lambda))\gamma t}$.} Making this replacement and using the spectral representation \eqref{022},
\begin{align*}
	\e^{\i t(\wt L_\S+\lambda^2\wt\Lambda)}=& |\wt\Omega_0\rangle \langle\wt\Omega_0|+  \sum_{j=1}^{d-1} \e^{\i t \lambda^2 \wt{\lambda}_{0,j}} \wt{Q}_{0,j} +\sum_{\widetilde e\not=0} \sum_{j=1}^{m_e} \e^{\i t (\wt{e}+\lambda^2 \wt{\lambda}_{\wt{e},j})} \wt{Q}_{\wt{e},j},
\end{align*}
we obtain
\begin{equation}
	\label{021}
\omega_0\big(\alpha^t_\lambda(X\otimes\bbbone_\R) \big) = \scalprod{JC^*C\wt\Omega_0}{\e^{\i t(\wt L_\S+\lambda^2\wt\Lambda)}(X\otimes\bbbone_\S\otimes\bbbone_\R)\wt\Omega_0} + R_{\lambda,t}(X),
\end{equation}
where $R_{\lambda,t}(X)$ satisfies \eqref{finalremainder}. According to the definition \eqref{012} of $\tau^t_\lambda$, the main term on the right side of \eqref{021} is 
\begin{equation*}
\scalprod{JC^*C\wt\Omega_0}{(\tau^t_\lambda(X)\otimes\bbbone_\s\otimes\bbbone_\R)\wt\Omega_0} = \omega_0(\tau^t_\lambda(X)).
\end{equation*}
This shows the representation \eqref{011}.

\subsection{Proof of complete positivity of $\tau^t_\lambda$. }
\label{cpsect}

We first show complete positivity of the weak coupling dynamics $\sigma^t_\lambda$, \eqref{08}. Then we modify that argument just slightly to show complete positivity of $\tau^t_\lambda$.

\subsubsection{Complete positivity of the weak coupling dynamics $\sigma^t_\lambda$.}
\label{cpwcsect}
Using \eqref{01} and a  density argument, one sees that for any system-reservoir state $\omega$,
	\begin{equation}
		\label{010}
	\lim_{\lambda\rightarrow 0} \omega\big(\alpha^{t/\lambda^2}_\lambda\circ\alpha_0^{-t/\lambda^2}(X\otimes\bbbone_\R)\big) = \omega\big(\bar\sigma^t(X)\otimes\bbbone_\R \big),
	\end{equation}
	where $\bar\sigma^t(X)$ is defined by $\big( \bar\sigma^t(X)\otimes\bbbone_\S \big) \Omega_{\S,\beta} = \e^{\i t \Lambda}\,  \big(X\otimes\bbbone_\S\big) \Omega_{\S,\beta}$ and satisfies (see \eqref{08}) $\sigma^t_\lambda = \bar\sigma^{\,\lambda^2t}\circ\alpha^t_\S =\alpha^t_\S\circ \bar\sigma^{\,\lambda^2t}$. Since  $\alpha^t_\S$ is completely positive, complete positivity of $\sigma^t_\lambda$ follows from that of  $\bar\sigma^t$. Let $\omega_{\R,\beta}$ be the reservoir equilibrium state and  let $P_\R$ be the partial trace over the reservoir, relative to  $\omega_{\R,\beta}$, defined by (linear extension of) $P_\R (X\otimes B) P_\R = X\,\omega_{\R,\beta}(B)$. Taking $\omega=\omega_\S\otimes\omega_{\R,\beta}$ in \eqref{010}, where $\omega_\s$ is any system state, gives
	$$
	\lim_{\lambda\rightarrow 0}\omega_\S\big( P_\R \alpha^{t/\lambda^2}_\lambda\circ \alpha_0^{-t/\lambda^2}(X\otimes\bbbone_\R) P_\R\big) = \omega_\S(\bar\sigma^t(X)).
	$$
	As the system is finite-dimensional, this is equivalent to
	$$
	\lim_{\lambda\rightarrow 0}P_\R \alpha^{t/\lambda^2}_\lambda\circ \alpha_0^{-t/\lambda^2}(X\otimes\bbbone_\R) P_\R = \bar\sigma^t(X).
	$$
	The left side is the limit of a family of completely positive maps. Hence $
	\bar\sigma^t$ is completely positive as well.

\subsubsection{Complete positivity of $\tau^t_\lambda$.} We denote by $\gamma^t_{\lambda,\mu}(\cdot) =\e^{\i t\wt L_\mu}\cdot \e^{-\i t\wt L_\mu}$ the dynamics of $\frak M$ generated by the Liouville operator $\wt L_\mu$ defined in \eqref{auxliouv}. The level shift operator of $\wt L_\mu$ is  $\lambda^2\mu^2\wt\Lambda$ (see also \eqref{016}). Repeating the argument of the weak coupling limit, we have (c.f. \eqref{010})
\begin{equation}
	\label{0010}
\lim_{\mu\rightarrow 0} \omega\big(\gamma_{\lambda,\mu}^{t/\mu^2}\circ\wt\gamma_{\lambda,0}^{-t/\mu^2}(X\otimes\bbbone_\R)\big) = \omega\big(\bar\tau_\lambda^t(X)\otimes\bbbone_\R \big),
\end{equation}
where $\bar\tau^t_\lambda$ is defined by $(\bar\tau^t_\lambda(X)\otimes\bbbone_\S)\wt\Omega_{\S,\beta,\lambda} = \e^{\i t\lambda^2\wt\Lambda}(X\otimes\bbbone_\S)\wt\Omega_{\S,\beta,\lambda}$.  Thus, by the same argument as in Section \ref{cpwcsect}, $\bar\tau^t_\lambda$ is completely positive, and hence so is $\tau^t_\lambda= \bar\tau^t_\lambda\circ \wt\alpha_{\S,\lambda}$, where $\wt\alpha_{\S,\lambda}(\cdot) =\e^{\i t\wt H_\S}\cdot\e^{-\i t\wt H_\S}$.

\bigskip
\bigskip

\noindent
{\bf Acknowledgement.\ } This work has been supported by an NSERC Discovery Grant and an NSERC Discovery Grant Accelerator.

\end{document}